\documentclass[12pt,a4paper]{article}

\usepackage{amssymb}
\usepackage{epsf}
\begin{document}

\title{\textbf{Lower bound on the electroweak wall velocity from hydrodynamic
instability}}
\author{
Ariel M\'{e}gevand\thanks{Member of CONICET, Argentina. E-mail address:
megevand@mdp.edu.ar},~ %
Federico Agust\'{\i}n Membiela\thanks{Member of CONICET, Argentina. E-mail
address: membiela@mdp.edu.ar}~
and Alejandro D. S\'anchez\thanks{%
Member of CONICET, Argentina. E-mail address: sanchez@mdp.edu.ar} \\[0.5cm]
\normalsize \it IFIMAR (CONICET-UNMdP)\\
\normalsize \it Departamento de F\'{\i}sica, Facultad de Ciencias Exactas
y Naturales, \\
\normalsize \it UNMdP, De\'{a}n Funes 3350, (7600) Mar del Plata, Argentina }
\date{}
\maketitle

\begin{abstract}
The subsonic expansion of bubbles in a strongly first-order electroweak phase
transition is a convenient scenario for electroweak baryogenesis. For most
extensions of the Standard Model, stationary subsonic solutions (i.e.,
deflagrations) exist for the propagation of phase transition fronts. However,
deflagrations are known to be hydrodynamically unstable for wall velocities
below a certain critical value. We calculate this critical velocity  for several extensions of the Standard Model and compare with an estimation of the wall velocity. In general, we find a region in parameter space which gives stable deflagrations as well as favorable conditions for electroweak baryogenesis.
\end{abstract}

\section{Introduction} \label{intro}

A first-order electroweak phase transition may explain the observed baryon asymmetry of the universe (BAU). Indeed, such a phase transition would
provide all the Sakharov conditions, namely, baryon number violation,
$C$ and $CP$ violation, and a departure from thermal equilibrium. For
a quantitatively successful electroweak baryogenesis (EWB) an extension of the
Standard Model (SM) is needed, such that there is enough $CP$ violation as well
as a sufficiently strong first-order phase transition.
In the standard mechanism for EWB  (see \cite{mr12} for a recent review), the departure from equilibrium acts in two different
ways. On the one hand, the expansion of bubbles of the broken-symmetry phase
builds up non-equilibrium particle densities in front of the bubble walls.  These densities are asymmetric for left handed particles and their antiparticles
due to $CP$ violating interactions with the wall.
This asymmetry is transported to the unbroken-symmetry phase, where it biases the
weak sphaleron processes which violate baryon number. The generated baryon asymmetry reenters the broken-symmetry phase. As a result, the bubble walls leave behind a net baryon number density.
On the other hand, before this baryon number density recovers the equilibrium, the baryon number violating processes should be turned off. Otherwise the generated BAU would be washed out. Such a suppression of the  sphaleron processes indeed occurs inside the bubble, as long as  the Higgs background field $\phi_{b}$ in the broken-symmetry phase satisfies
the well known condition
\begin{equation}
{\phi_{b}}/{T}\gtrsim1, \label{washout}
\end{equation}
where $T$ is the temperature. The ratio $\phi_{b}/T$ plays
the role of an order parameter, and the condition (\ref{washout}) expresses
the baryogenesis requirement of a strongly first-order phase transition.

Although for a Higgs mass as large as $125\mathrm{GeV}$ the electroweak phase transition is a
smooth crossover, many extensions of the SM give strongly first-order phase
transitions. Most investigations of EWB concentrate in the value of the order parameter $\phi_{b}/T$ and
the sources of CP violation for specific models.
Since the computation of the
velocity $v_w$  of bubble walls is too involved, a specific value is often assumed (typically
$v_{w}=0.1$) to obtain a result for the BAU\footnote{It is worth mentioning that, in contrast, for the generation of gravitational waves higher
velocities are preferable, since the collisions of faster walls produce gravitational waves of higher intensity (see, e.g., \cite{kkt94}). As a consequence, a supersonic wall velocity is generally assumed.}.
However, the generated BAU has also an important
dependence on $v_{w}$. Indeed, for very small
velocities thermal equilibrium is restored and a small baryon asymmetry
is generated. On the other hand, if the wall velocity is too large the diffusion of
left-handed density perturbations is not efficient, and
the resulting baryon number density is again small. In other words,
a departure from equilibrium is needed, but such a departure should not be too strong. As a consequence, the generated baryon asymmetry has a maximum for
a certain wall velocity $v_{w}=v_{\mathrm{peak}}$.  The value
of $v_{\mathrm{peak}}$ depends on the time
scales associated to particle diffusion and baryon number violation, and is in general in the range
$10^{-2}<v_{\mathrm{peak}}<10^{-1}$ (see, e.g., \cite{ck00,cjk00,cmqsw01}).
A sizeable BAU is more easily obtained if $v_w$ is close to $v_{\mathrm{peak}}$. Moreover, any
model which gives supersonic velocities is in conflict with the standard EWB mechanism \cite{jpt96}.

Subsonic wall velocities are possible  due to the friction with the plasma, which generally causes
the walls to reach a terminal velocity.
This velocity is given by the balance between the driving force, which depends
on the pressure in the two phases, and the friction force,
which depends on the microscopic interactions of the bubble wall with plasma
particles. The driving force is very sensitive to  hydrodynamics. As a
consequence of non-linear hydrodynamics, there are different kinds of stationary solutions for the propagation of the wall \cite{landau}. The solutions which can be realized in a cosmological phase transition (see, e.g.,
\cite{gkkm84,eikr92,k85,kk86,ikkl94,kl95,kl96}) are \emph{weak deflagrations}, which are subsonic,  \emph{Jouguet deflagrations}, which are
supersonic, and \emph{Jouguet or weak detonations}, which are supersonic too. Hence, the case of interest for baryogenesis is that of weak deflagrations.

It is well known that the stationary propagation of a weak deflagration front may be unstable \cite{landau}. For the case of a relativistic equation of state, the
stability of deflagrations was first studied in Ref.~\cite{link}. The result
was that deflagrations are always unstable under perturbations above a certain
wavelength. This analysis was improved in Ref.~\cite{hkllm}. The
main improvement was to take into account the dependence of the stationary
velocity on the temperature. The main result of Ref.~\cite{hkllm} was that the
deflagration is stable for wall velocities above a certain critical value
$v_{\mathrm{crit}}$. Numerical simulations \cite{fa03} agree with such a
stabilization. In Ref.~\cite{mm14defla}, the results of \cite{hkllm} were
improved by taking into account temperature fluctuations on both sides of the
wall, as well as the fact that the reheating due to the release of latent heat depends
on the wall velocity. For
small amounts of supercooling, the results of Ref.~\cite{mm14defla} agree with
those of Ref.~\cite{hkllm}.

In Ref.~\cite{hkllm}, the stability analysis was applied to
the electroweak phase transition for the minimal Standard Model with
unrealistic values of the Higgs mass, which gives a strong enough phase
transition for EWB. For Higgs masses higher than
$m_{H}=40GeV$, the critical velocity below which deflagrations are unstable
was found to be $v_{\mathrm{crit}}\lesssim0.07$. This result was compared with the wall velocity calculations \cite{dlhll92,lmt92}, which gave $v_{w}\gtrsim0.1$. Therefore, the result
of Ref.~\cite{hkllm} indicated that the electroweak deflagration is stable.
However, both $v_{w}$ and $v_{\mathrm{crit}}$ depend on the model and should
be recalculated for each extension of the SM.

The calculation of $v_{w}$ is more involved than that of $v_{\mathrm{crit}}$
and depends on more details of the model. Indeed, the value of the critical
velocity depends only on thermodynamical parameters which can be derived from
the free energy density. In contrast, the actual value of the stationary
velocity depends (besides thermodynamics and hydrodynamics) on the friction of
the wall with the plasma. The computation of the friction force involves
considering Boltzmann equations for the out-of-equilibrium particle densities
in front of the wall. For the SM, a thorough calculation
(including reheating effects) \cite{mp95b} gave wall velocities in the range
$0.36<v_{w}<0.44$ (for $0<m_{H}<90\mathrm{GeV}$). A similar calculation for the Minimal Supersymmetric Standard Model (MSSM) \cite{js01} gave smaller velocities, $v_{w}=(5-10)\times10^{-2}$, due to the larger
particle content of this model (essentially, due to the contribution of top squarks). To our knowledge, these two results constitute the only detailed microphysics
calculations for specific models. The reason for this is
the difficulty of computing the collision terms for the Boltzmann equations. In
spite of this, many investigations of the friction were performed. In
particular, a study of the overdamped evolution of gauge fields \cite{m00}
showed that infrared boson excitations generally increase the friction and,
consequently, cause smaller wall velocities than previous studies. In
particular, for the SM the estimated wall velocity was $v_{w}\lesssim0.01$ for
$m_{H}\simeq80\mathrm{GeV}$ and $v_{w}\simeq0.1$ for
$m_{H}\simeq45\mathrm{GeV}$.

In this paper we shall investigate the possible instability of the electroweak
wall velocity for several extensions of the SM. The main motivation for this is the fact that the deflagration instability may affect the baryogenesis
scenario. Indeed, notice that the value of $v_{\mathrm{crit}}$ obtained in
Ref.~\cite{hkllm} lies within the optimal range for EWB.
Moreover, given the general uncertainties and large errors in the estimations of $v_{w}$, the value of $v_{\mathrm{crit}}$, which is much easier
to calculate, provides a lower bound for $v_{w}$ which may be important to
constrain baryogenesis. It is worth mentioning also that an instability of the stationary wall
propagation may have several cosmological consequences, such as the generation
of magnetic fields \cite{soj97} or gravitational waves \cite{mm14defla}.

The plan of the paper is the following.
In  Sec.~\ref{stat} we review the
hydrodynamics of a wall which propagates as a deflagration and we discuss the stability of such a stationary solution as a function of
thermodynamic parameters. In Sec.~\ref{results} we calculate the critical
velocity below which the wall becomes unstable. We consider the electroweak
phase transition for several extensions of the Standard Model. We also
estimate the wall velocity for each model  in order to study the stability
as a function of the parameters. In Sec.~\ref{discu} we discuss on the possible
consequences of the instability. Finally, in Sec.~\ref{conclu} we summarize our
conclusions.
Details of the calculation of the phase transition dynamics are contained in App.~\ref{apdyn}.
Further discussion on the critical velocity as well as a fit can be found in App.~\ref{apvcrit}.

\section{Stationary wall propagation and hydrodynamic stability} \label{stat}

\subsection{First-order electroweak phase transition}

The relevant quantity describing the phase transition is the free energy density or finite-temperature effective potential\footnote{In this work we shall only consider models which can be described (at least, approximately)
by a single Higgs field $\phi$.} $\mathcal{F}(\phi,T)$. At a given temperature $T$, the minima of $\mathcal{F}$ give the possible thermal expectation values of the Higgs field $\phi$, which determine the different phases. For the electroweak theory, we have a phase transition from the symmetric phase to the broken-symmetry phase at a critical temperature $T_c\sim 100$GeV. In the minimal SM, the electroweak phase transition is just a smooth crossover. In Sec. \ref{results} we shall consider several extensions of the SM for which the electroweak phase transition is first-order.

For a first-order phase transition, there is a range of temperatures around
$T_{c}$ for which the effective potential has two minima separated by a
barrier. For $T>T_{c}$ the symmetric minimum $\phi=0$ is the absolute minimum
of $\mathcal{F}(\phi,T)$, while for $T<T_{c}$ the absolute minimum has a
nonvanishing value $\phi_{b}(T)$. We shall use subindexes $u$ and $b$ for the
unbroken- and broken-symmetry phase, respectively. These phases are thus
characterized by the free energy densities $\mathcal{F}_{u}(T)=\mathcal{F}
(0,T)$ and $\mathcal{F}_{b}(T)=\mathcal{F}(\phi_{b}(T),T)$. The critical
temperature is  given by the equation $\mathcal{F}_{u}(T_{c}
)=\mathcal{F}_{b}(T_{c})$. The energy density and pressure for each phase are obtained from
the free energy density through $\rho(T)=\mathcal{F}(T)-T\mathcal{F}^{\prime
}(T),p(T)=-\mathcal{F}(T)$, where a prime indicates a derivative with respect to the temperature.

A  first-order phase transition occurs via the nucleation and expansion of bubbles.
As we shall see, the relevant parameters for our calculation will be
the latent heat $L$ defined as $L\equiv\rho_{u}\left(  T_{c}\right)
-\rho_{b}\left(  T_{c}\right)  $, the  enthalpy density  before the phase transition, $w_u(T_c)=\rho_u(T_c)+p_u(T_c)$, and  the nucleation temperature $T_n$. The latter is the temperature at which bubbles effectively begin to nucleate. We describe its calculation in App. \ref{apdyn}. The enthalpy density and latent heat are given by
\begin{eqnarray}
w_{u}(T_{c})&=&-T_{c}\mathcal{F}_{u}^{\prime}(T_{c}), \label{dobleve}
\\
L&=&T_{c}\left[  \mathcal{F}_{b}^{\prime}(T_{c})-\mathcal{F}_{u}^{\prime}
(T_{c})\right]  . \label{ele}
\end{eqnarray}

After nucleating, bubbles expand due to the higher pressure of the stable
phase. In most cases, the bubble walls quickly reach a terminal velocity due
to the friction with the plasma. We shall
concentrate in such a case. From the time a bubble nucleates to the time their
walls collide with other bubbles, the temperature of the plasma varies due to
the adiabatic expansion of the universe and due to the release of latent heat.
As a consequence, the wall velocity will vary too. We shall use $T=T_{n}$ as a
representative value for the temperature of the phase transition.

\subsection{Microphysics and hydrodynamics}

The motion of a bubble wall can be derived from
the equation for the Higgs
field  in the plasma  (see, e.g.,
\cite{dlhll92,lmt92,mp95b,js01,m00,t92,k92,a93,mp95a}),
\begin{equation}
\partial _{\mu }\partial ^{\mu }\phi +\frac{\partial \mathcal{F}( \phi
,T) }{\partial \phi }
+\sum_i
\frac{dm_{i}^{2}}{ d\phi }\int \frac{d^{3}p}{( 2\pi) ^{3}2E_{i}}\delta f_{i}
 =0, \label{eqphi}
\end{equation}%
with $E_i=\sqrt{\mathbf{p}^2+m_i^2}$, where $m_i$ are the $\phi$-dependent particle masses, $\delta f_i$ are the  deviations  of particle densities from equilibrium, and the sum runs over all particle species.
The last term gives the friction with the plasma. In order to transform Eq. (\ref{eqphi}) into an equation for the bubble wall, the general procedure is to use some approximation or ansatz  for the field profile (which is static in the reference frame of the wall) and integrate across the wall.
Since the deviations $\delta f_i$ depend on the wall velocity, the last term  in (\ref{eqphi}) gives the friction force $F_{\mathrm{fr}}$, while the second term gives the driving force $F_{\mathrm{dr}}$. Thus,
the steady state velocity of a bubble wall is given by the force balance $F_{\mathrm{dr}}=F_{\mathrm{fr}}$ (for a more detailed explanation, see e.g.~\cite{ariel13}).

The driving force is relatively easy to calculate.    In particular, if the temperature remains constant across the wall, we have $F_{\mathrm{dr}}=p_{b}(T)-p_{u}(T)$. On the other hand,
the friction is proportional to the departure of the plasma particles from
their equilibrium distributions. This departure from equilibrium depends not
only on the interaction of the particles with the Higgs field at the wall, but
also on the interactions of plasma particles away from the wall. Thus, the
calculation involves solving a system of Boltzmann equations for the
population densities of the relevant species. The Boltzmann equations include
collision terms which must be computed by calculating the scattering rates for
all the relevant processes.
Such a calculation is often referred to as the \emph{microphysics} calculation.

As already mentioned in Sec.~\ref{intro}, the microphysics calculation is very difficult. In particular, the computation of the collision terms was carried out only for a couple of models and in the non-relativistic (NR) case \cite{mp95b,js01,m00}.
The result is a friction force of the form $F_{\mathrm{fr}}=\eta_{\mathrm{NR}} v_{w}$.
The friction coefficient $\eta_{\mathrm{NR}}$ is very model dependent, and its calculation involves the use of several  approximations.
The ultra-relativistic (UR)
limit has also been considered \cite{bm09}. It turns out that this limit is even simpler than the NR case. The result is
that the friction  saturates for $v_w\to 1$, i.e., the friction force
reaches a velocity-independent value $F_{\mathrm{fr}}=\eta_{\mathrm{UR}}$.
Intermediate cases are much more difficult to treat. In a recent treatment \cite{knr14}, the friction was considered beyond the small wall velocity regime. However, the deviations from equilibrium were still considered to be small. In particular, the friction force calculated in Ref.~\cite{knr14} does not match the UR results of Ref.~\cite{bm09}.

In order to overcome the difficulties of the microphysics calculation, a phenomenological approach
has often been used, which consists in replacing the last term in Eq.~(\ref{eqphi}) with an effective damping term of the form
$u^{\mu }\partial _{\mu }\phi$, where $u^\mu=(\gamma,\gamma\mathbf{v})$ is the four velocity of the fluid (see, e.g., \cite{ikkl94,ms12,ms09}).
If we
ignore hydrodynamics, this approach gives a friction force of the form $F_{\mathrm{fr}}=\eta v_w$ in the NR limit, where
the coefficient $\eta$  is a free parameter coming from the phenomenological
damping term. Hence, setting $\eta$ to the value $\eta_{\mathrm{NR}}$ from the microphysics calculation gives the correct friction force in this limit.
This phenomenological model extrapolates the
NR behavior $\eta v_w$ to
a force of the form $\eta\gamma v$
for
relativistic velocities,
where $v$
is the velocity of the fluid relative to the wall (which will vary across the wall, see below). However, this simple model does not give the friction saturation in the UR limit. In order to reproduce the saturating behavior, in Ref.~\cite{ekns10} a phenomenological model which gives a friction force of the form $\eta v$ was considered. However, such a model with a single free parameter $\eta$ can hardly match both the NR and UR forces $\eta_{\mathrm{NR}}v$ and $\eta_{\mathrm{UR}}$. In Ref.~\cite{ariel13} a
phenomenological interpolation between the two regimes was considered.

For the subsonic velocities we are interested in, the exact dependence of the friction on the velocity does not introduce significant effects \cite{ariel13}. Therefore, we shall use the phenomenological scaling $\eta v\gamma$,
which will allow us to use the results of the stability analysis \cite{mm14defla}. We shall discuss the implications of this choice in Sec.~\ref{discu}.
In order to estimate the wall velocity for specific models, in Sec. \ref{results} we shall set the parameter $\eta$ to the value
$\eta_{\mathrm{NR}}$ coming from microphysics calculations.

Using this phenomenological approach to the friction, it is not difficult to include the hydrodynamics, i.e., to take into account the change of
fluid variables across the wall. The fluid variables in each phase are related
by matching conditions at the phase discontinuity. In the rest frame of the
wall, we have \cite{landau}
\begin{eqnarray}
w_{u}\gamma_{u}^{2}v_{u}  &  =&w_{b}\gamma_{b}^{2}v_{b},\nonumber \\
w_{u}\gamma_{u}^{2}v_{u}^{2}+p_{u}  &  =&w_{b}\gamma_{b}^{2}v_{b}^{2}
+p_{b}, \label{eqlandau} \\
\mathbf{v}_{u}^{\perp}  &  =&\mathbf{v}_{b}^{\perp},\nonumber
\end{eqnarray}
where $v$ is the component of the fluid velocity along the wall motion,
$\mathbf{v}^{\perp}$ is the velocity in the transverse direction, and
$\gamma=1/\sqrt{1-\mathbf{v}^{2}}$. By symmetry, we set $\mathbf{v}^{\perp
}=\mathbf{0}$ for the stationary motion, but this component must be taken into account in the stability analysis. Furthermore, we define the incoming
and outgoing flow velocities by $-v_{u}$ and $-v_{b}$, respectively, so that
we deal with positive values of the variables $v_{u},v_{b}$. Using suitable approximations (see App. \ref{apdyn}), one obtains a  friction force
\begin{equation}
 F_{\mathrm{fr}}=
\eta\left\langle \gamma v\right\rangle , \label{ffr}
\end{equation}
where  $\langle f\rangle=\frac{1}{2}(f_{b}+f_{u})$ for a quantity $f$
defined on each side of the wall.

The  thermodynamical quantities $w(T),p(T)$ in Eqs.~(\ref{eqlandau}) are related by the equation of state.
The treatment of hydrodynamics is considerably simplified by considering a
simple approximation for the equation of state. This is particularly important
for the stability analysis. In order to use the analytical results of
Ref.~\cite{mm14defla},  we shall consider the bag EOS.
This is the
simplest EOS which can describe a phase transition\footnote{One
limitation of this model is the fact that it gives a constant speed of sound
$c_{s}=1/\sqrt{3}$ in both phases, while the actual value of $c_{s}$ may
depart from this value \cite{lm14}.}. Due to the difficulty of the stability analysis, it is not
trivial to generalize the results of Ref.~\cite{mm14defla} beyond these
approximations.
With these approximations, the driving  force  becomes
\begin{equation}
F_{\mathrm{dr}}=\frac{L}{4}\left(  1-\frac{T_{u}^{2}T_{b}^{2}}{T_{c}^{4}}\right).  \label{fdr}
\end{equation}

Several of the quantities in Eqs.~(\ref{ffr}-\ref{fdr}) are related through Eqs.~(\ref{eqlandau}) and boundary conditions (see App.~\ref{apdyn}). As a result, the wall velocity $v_w$ depends only on  the temperature $T_u$, the coefficient $\eta$, and the parameter
\begin{equation}
\bar{L}\equiv{L}/{w_{u}(T_{c})}. \label{lbar}
\end{equation}
Notice that in general the driving force does
not coincide with the pressure difference $p_{b}-p_{u}$, as one might expect, since it is nontrivially affected by hydrodynamics.
Nevertheless, the fact that  the pressure difference changes sign at $T=T_{c}$
is reflected in
$F_{\mathrm{dr}}$, which
is very sensitive to the departure from
the critical temperature.
In particular,  the reheating of the fluid causes a decrease of $F_{\mathrm{dr}}$, acting as
an effective  friction \cite{ms09,kn11}.

There are in general several solutions for $v_w$. The only solution for which the
wall velocity is subsonic  is the weak deflagration, and we shall only be
interested in this case. Thus, in the bag approximation, we have $v_w<1/\sqrt{3}$. The deflagration solution exists for large
enough friction and small enough supercooling. For a deflagration, the fluid is reheated in front of the wall. Therefore, the temperature $T_u$ is higher than the nucleation temperature $T_n$. The relation between $T_u$ and $T_n$ introduces an equation to solve together with that corresponding to the equilibrium  of the forces (\ref{ffr}-\ref{fdr}). We write down all these equations in App.~\ref{apdyn}.

\subsection{Stability of the deflagration}

The possible hydrodynamic instabilities of cosmological phase-transition fronts
have been investigated in
Refs.~\cite{link,hkllm,fa03,mm14defla,abney,r96,mm14deto}. The linear stability
analysis of the wall-fluid configuration consists of considering small
perturbations of the fluid variables on both sides of the wall, together with
small  deformations of the planar and infinitely thin wall. Below we briefly
sketch the generalities of the calculation for the deflagration case. For
detailed and more general analysis, see \cite{mm14defla,mm14deto}.

There are in principle seven variables, namely, the wall deformation $\zeta$,
the pressure fluctuations $\delta p_{u},\delta p_{b}$, the variations of the
fluid velocity along the propagation direction $\delta v_{u},\delta v_{b}$, and
the transverse velocities $v_{u}^{\perp},v_{b}^{\perp}$. These perturbations
depend on space and time. The three fluid fluctuations on a given side of the
wall are related by the fluid equations. Linearizing these equations and
looking for solutions of the form $\exp(i\mathbf{k}
\cdot\mathbf{x}^{\perp}+qz+\Omega t)$, one obtains dispersion relations for
$q(\mathbf{k,}\Omega)$, as well as algebraic equations relating the amplitudes
of the different fluctuations. For the stability analysis one is interested in
the unstable modes, which correspond to $\Omega>0$ and $qz<0$ \cite{landau}.
For weak deflagrations, we have one unstable mode in front of the wall and two
unstable modes behind it. We are thus left with four unknowns, namely, the
amplitudes of the three unstable fluid modes and that of the wall perturbation.
Finally, the fluid perturbations on the two sides of the wall are related by
junction conditions at the wall, which are the counterparts of
Eqs.~(\ref{eqlandau}). There is also an equation for the perturbations of the
surface, which is the counterpart of  the equation
$F_{\mathrm{dr}}=F_{\mathrm{fr}}$. As a consequence, one obtains a system of
four algebraic equations for the four unknowns. The weak deflagration is
linearly unstable if a nontrivial solution exists for this system.

Looking for linear instability is thus equivalent to demanding a $4\times4$
determinant to vanish. This gives an equation for the  growth rate $\Omega$ as
a function of the perturbation wavenumber $k\equiv|\mathbf{k}|$. For a complete
treatment and analytic expressions we refer the reader to
Ref.~\cite{mm14defla}. The general result is that the deflagration is unstable
for wavenumbers below a certain value $k_{c}$ which depends on the
thermodynamic parameters and the wall velocity. Nevertheless, depending on the
parameters the critical wavenumber $k_{c}$ may become negative, in which case
all wavenumbers are stable. Demanding this to be the case (i.e., $k_{c}\leq0$)
one obtains a condition for the wall velocity, namely, $v_{w}\geq
v_{\mathrm{crit}}$ for  stability of the deflagration. The critical velocity
$v_{\mathrm{crit}}$ corresponds to $k_{c}=0$, which is equivalent to the
equation\footnote{There are some sign differences with respect to
Ref.~\cite{mm14defla} because we have defined $v\equiv|v|$ for the fluid
velocity in the wall frame.}
\begin{equation}
\Delta v\left[  \gamma_{sb}^{2}\gamma_{b}(1-\beta_{b})-\frac{\gamma_{b}}
{2}\right]  +\frac{v_{u}}{\gamma_{b}^{2}}\left\langle \gamma_{s}^{2}
\gamma(1-\beta)\right\rangle =0, \label{eqvcrit}
\end{equation}
where $\gamma_{s}^{2}=1-v^{2}/c_{s}^{2}$, $\Delta
v=v_{b}-v_{u}$. For the bag model, the quantity $\beta$ is
given by
\begin{equation}
\beta_{u,b}=\frac{4\langle\gamma v\rangle v_{u,b}}{\gamma_{u,b}\left(
\frac{T_{c}^{4}}{T_{u}^{2}T_{b}^{2}}-1\right)  }. \label{betabag}
\end{equation}
For $v_w<v_{\mathrm{crit}}$, perturbations with wavenumbers $0<k<k_c$ grow
exponentially. As discussed in Sec.~\ref{discu} below, for the electroweak
phase transition the characteristic time for the development of the
instabilities is generally much shorter than the time scales associated to
bubble growth, even if $v_w$ is very close to $v_{\mathrm{crit}}$.

It is worth remarking that the critical value  of the wall velocity does not
depend on the friction coefficient $\eta$. The parameter $\eta$ determines  the
\emph{actual} value of $v_w$. The critical velocity $v_\mathrm{crit}$ can thus
be associated to a critical friction $\eta_\mathrm{crit}$ (which could then be
compared with the actual value of $\eta$ in order to determine the stability).
However, \emph{the form} of the friction force  (\ref{ffr}) is implicit in the
equations above. This is because, in the stability analysis, the coefficient
$\eta$ is eliminated by writing it as a function of the velocity and the
driving force \cite{hkllm,mm14defla}.

\begin{figure}[bth]
\centering
\epsfysize=6cm \leavevmode \epsfbox{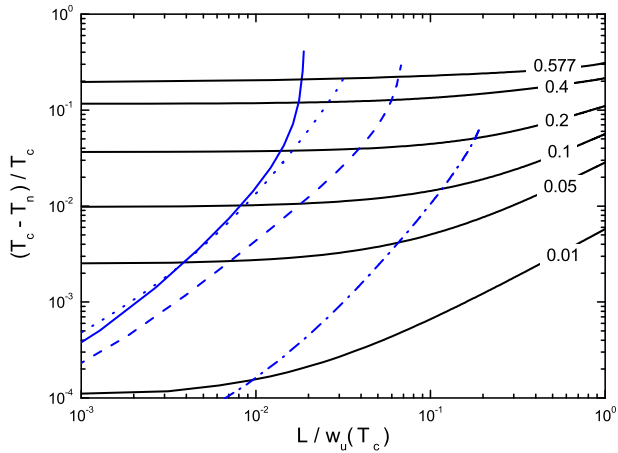}
\caption{Curves of constant $v_{\mathrm{crit}}$ (solid black lines) in the plane of the parameters
$\bar{L}$ and $1-T_n/T_c$. Blue curves correspond to extensions of the SM considered below. Solid line:
strongly
coupled bosons (2 d.o.f.). Dashed line: strongly coupled bosons (12 d.o.f.).
Dash-dotted line: strongly coupled fermions. Dotted line: dimension-six operator.}
\label{figspace}
\end{figure}

The critical velocity depends only on the dimensionless parameters $\bar{L}$
and $T_{n}/T_{c}$. Both parameters range between $0$ and $1$ and quantify the
amounts of latent heat and supercooling, respectively. In Fig.~\ref{figspace}
we show the curves of constant $v_{\mathrm{crit}}$ in the space of these two
parameters [we considered the parameter $(T_{c}-T_{n})/T_{c}$ since in many
physical cases the temperature is very close to $T_{c}$]. These curves are
model independent. We also show the points in this plane corresponding to some
of the specific models considered below (those which span larger regions of the
plane). The limit $\bar{L}\rightarrow 0$, $T_{c}-T_{n}\rightarrow 0$
corresponds to a second-order phase transition. The critical velocity increases
with the amount of supercooling but is rather insensitive to the latent heat.
The curves of constant $v_{\mathrm{crit}}$ accumulate at the weak deflagration
limit $v_{\mathrm{crit}} =c_{s}\simeq0.577$. For values of the parameters above
this curve, there is no critical velocity and any subsonic velocity is
unstable. In App. \ref{apvcrit}  we consider in more detail the dependence of
$v_{\mathrm{crit}}$ on these parameters and we give a simple fit for
$v_{\mathrm{crit}}(\bar{L},T_n/T_c)$.

In Fig.~\ref{figspace} it seems that, for physical models, none of the two parameters
$\bar{L}$ or $(T_{c}-T_{n})/T_{c}$ can reach the upper value 1.
In fact, the supercooling parameter may in principle take values arbitrarily
close to $1$, since the nucleation temperature $T_{n}$ can be very small for
models with barriers in the zero-temperature effective potential. The upper
limits of $(T_{c}-T_{n})/T_{c}$ in these curves are due to the break-down of
our calculations for such strong phase transitions. On the other hand, it is true that the latent heat tends to take relatively small values $\bar
{L}\lesssim0.1$ for
physical models, as seen in the figure. This is because, in general, the
entropy released in the phase transition is only a fraction of the total
entropy. This fraction is proportional to the fraction of degrees of freedom (d.o.f.)~which are strongly coupled to the Higgs.

As can be seen in Fig.~\ref{figspace}, the wide range of possible amounts of supercooling
implies a wide range of possible values of
the critical velocity $v_{\mathrm{crit}}$. On the other hand, the actual value of the wall
velocity, $v_w$, also grows with the amount of supercooling, and it is not
easy to guess, without specific calculations, in which cases the deflagration
will be unstable. We perform such calculations in the next section.

\section{Electroweak phase transition models \label{results}}

Several extensions of the SM  have been considered in the
literature in order to increase the strength of the phase transition and obtain a sizeable electroweak baryogenesis.
Constraints from the recent LHC results threaten the viability of
baryogenesis in some models (see, e.g., \cite{clw13,cnqw13,kp14}). Since our aim is a general investigation of the possible instability, we shall not discuss the implications of experimental
constraints. Our results will give in principle an additional constraint for
the baryogenesis scenario. Furthermore, we shall not limit ourselves to the case of strongly first-order
phase transitions, since the instability of the stationary wall propagation
may have cosmological consequences even for weakly first-order phase
transitions.

A classification of models for the electroweak phase transition was recently given in Ref.~\cite{clw13}.
Besides these model classes, we shall consider a model with TeV fermions
introduced in Ref.~\cite{cmqw05} as well as two-loop effects. For simplicity, we shall consider a single background field $\phi$. For several
extensions of the SM,  $\phi$ corresponds to the SM Higgs,
$\langle H^{0}\rangle\equiv\phi/\sqrt{2}$. In extensions in which more than
one scalar develop a vacuum expectation value (VEV), it is sometimes a good
approximation to consider a single field $\phi$, corresponding to a certain
trajectory in field space.

\subsection{Free energy and friction}

We shall consider the one-loop finite-temperature effective
potential given by
\begin{equation}
\mathcal{F}(\phi,T)=V_{0}(  \phi)  +V_{1}(  \phi)
+\mathcal{F}_{1}(\phi,T), \label{ftot}
\end{equation}
where $V_{0}$ is the tree-level potential and $V_{1}$, $\mathcal{F}_{1}$ are
the zero-temperature and finite-temperature parts of the one-loop correction.
These corrections receive contributions from the SM particles and from
beyond-SM particles. We shall consider a spontaneous symmetry-breaking
potential of the form
\begin{equation}
V_{0}(  \phi)  =-m^{2}\phi^{2}+\frac{\lambda}{4}\phi^{4}
+\frac{\lambda}{4}v^{4} \label{v0}
\end{equation}
as well as some tree-level modifications. Here, the
parameters $m$ and $\lambda$ are related to the Higgs mass and VEV by $v=\sqrt{2/\lambda}m=246\mathrm{GeV}$, $m_{H}
=\sqrt{2\lambda v^{2}}=125\mathrm{GeV}$, and the constant term in Eq.~(\ref{v0}) was added so that the potential vanishes at the minimum, i.e., $V(v)=0$.
We shall assume Higgs-dependent masses of the form
\begin{equation}
m_{i}^{2}(\phi)=h_{i}^{2}\phi^{2}+\mu_{i}^{2}, \label{masses}
\end{equation}
and we shall consider the renormalized one-loop zero-temperature correction
\begin{equation}
V_{1}(  \phi)  =\sum_{i}\frac{\pm g_{i}}{64\pi^{2}}\,\left[
m_{i}^{4}(\phi)\left(  \log\frac{m_{i}^{2}(\phi)}{m_{i}^{2}(v)}-
\frac{3}{2}\right)  +2m_{i}^{2}(\phi)m_{i}^{2}(v)-\frac{m_{i}^{4}(v)}{2}\right]  ,
\label{v1loop}
\end{equation}
where the upper and lower signs correspond to bosons and fermions,
respectively, and $g_{i}$ is the number of d.o.f.~of particle species $i$. This
expression corresponds to the renormalization conditions that the tree-level
values of the minimum and Higgs mass are not shifted by radiative corrections, i.e.,
$V_{1}^{\prime}(v)=0$, $V_{1}^{\prime\prime}(v)=0$ (where a prime indicates a
derivative with respect to $\phi$). We have added a term $m_{i}^{4}(v)/2$ to
the well known expression, so that the true vacuum energy density is not
shifted either, i.e., $V_{1}(v)=0$. Thus, in the symmetric phase we will have a false
vacuum energy density given by $\rho_{\mathrm{vac}} =V_{0}\left(  0\right)
+V_{1}(0)$, which contributes to the Hubble rate during the phase transition.
Finally, the one-loop finite-temperature correction, including the resummed
daisy diagrams, is given by \cite{quiros}
\begin{eqnarray}
\mathcal{F}_{1}(\phi,T)  &  =&\sum_{i}\pm\frac{g_{i}T^{4}}{2\pi^{2}}\int
_{0}^{\infty}dx\,x^{2}\log\left[  1\mp\exp\left(  -\sqrt{x^{2}+m_{i}
^{2}(  \phi)  /T^{2}}\right)  \right] \nonumber\\
&  +&\sum_{bosons}\frac{g_{i}T}{12\pi}\left[  m_{i}^{3}(  \phi)
-\mathcal{M}_{i}^{3}(  \phi)  \right]  . \label{f1loop}
\end{eqnarray}
The last term in Eq.~(\ref{f1loop}) receives contributions from all the bosonic
species. We have $\mathcal{M}_{i}^{2}(  \phi)  =m_{i} ^{2}(
\phi)  +\Pi_{i}(  T)  $, where $\Pi_{i}( T)  $ are
the thermal masses. For the transverse polarizations of the gauge bosons we
have $\Pi(T)=0$ and the last term in (\ref{f1loop}) vanishes.

In general, Eqs.~(\ref{ftot}-\ref{f1loop}) give a phase transition at a
certain temperature $T_{c}$ from the high-temperature symmetric phase to the
low-temperature broken-symmetry phase. The
dynamics of the phase transition depends mostly on the difference
\begin{equation}
V_{T}(\phi)\equiv\mathcal{F}(\phi,T)-\mathcal{F}(0,T). \label{vt}
\end{equation}
Depending on the particle content of the model, the phase transition may be first order, i.e., the effective potential may have two minima separated by a barrier.
This is better appreciated in the
high-temperature approximation, i.e., when Eq.~(\ref{f1loop}) admits an
expansion in a power series of $m/T$. For masses of the form (\ref{masses}) we
obtain the well known simple form \cite{ah92}
\begin{equation}
V_{T}(\phi)=D\left(  T^{2}-T_{0}^{2}\right)  \phi^{2}-ET\phi^{3}+\frac
{\lambda_{T}}{4}\phi^{4}, \label{fpots}
\end{equation}
with coefficients given approximately by
\begin{equation}
D=\sum_{i}\tilde{c}_{i}\frac{g_{i}h_{i}^{2}}{24},\ T_{0}^{2}=\frac{1}{D}
\frac{m_{h}^{2}}{4},\ E=\sum_{\mathrm{t.g.b.}}\frac{g_{i}h_{i}^{3}}{12\pi
},\ \lambda_{T}=\lambda, \label{coefs}
\end{equation}
with $\tilde{c}_{i}=1$ for bosons and $1/2$ for fermions\footnote{Besides, we have $\mathcal{F}(0,T)\simeq\rho_{\mathrm{vac}}-\frac{\pi^{2}
}{90}g_{\ast}T^{4}$, where $g_{\ast}=\sum_{i}c_{i}g_{i}$, with $c_{i}=1$ for
bosons and $7/8$ for fermions. For the SM, we have $g_{\ast}\simeq107$.}. We have neglected for simplicity corrections to these coefficients which are suppressed by factors of order $g_{i}/(32\pi^{2})$. The terms $m^{4}\log m^{2}$ in (\ref{v1loop}) cancel out with similar terms in the expansion of the thermal integrals in Eq.~(\ref{f1loop}), and only a
dependence of the form $\phi^{4}\log T^{2}$ remains. This gives a soft
dependence of the coefficient $\lambda_{T}$ on $T$, which was neglected in the
approximation (\ref{coefs}). For the coefficient $E$ the sum runs only over
transverse gauge bosons. Indeed,  only the bosonic thermal integral  has a cubic
term $\sim(m/T)^{3}$ in its power expansion. The last term in Eq.~(\ref{f1loop}) replaces $m^{3}$
with $\mathcal{M}^{3}$ and we actually have
\begin{equation}
-\frac{gT}{12\pi}\left[  h^{2}\phi^{2}+\mu^{2}+\Pi(  T)  \right]
^{3/2} . \label{debye}
\end{equation}
For gauge bosons we have $\mu=0$, and for their transverse polarizations we also have $\Pi=0$. Hence, only in this case we obtain a contribution to the term $-ET\phi^{3}$.

It is well known that, for an effective potential of the form (\ref{fpots}), it is
precisely the cubic term the one which allows a first-order phase transition,
while for $E=0$ the transition is second order. Indeed, the value of the
order parameter at the critical temperature is given by $\phi_b(T_{c})
/T_{c}=2E/\lambda_{T_{c}}$. For the minimal SM, the constant $E$ is very small
and Eq.~(\ref{fpots}) gives a very weakly first-order phase transition\footnote{Moreover, the small
value of $\phi_b/T$ causes perturbation theory to break down and only
non-perturbative calculations become reliable.}.
The relevant SM contributions to the
one-loop effective potential come from the $Z$ and $W$ bosons, the top quark,
and the Higgs and Goldstone bosons. The Higgs sector is usually ignored in
the one-loop radiative corrections. This should be a good approximation in
extensions of the SM which include particles with strong couplings to $\phi$. We shall use this simplification.
Therefore, the $\phi$-dependent masses
are of the form $h_{i}\phi$, with $h_{i}=m_{i}/v$, where $m_{i}$ are the
physical masses at zero temperature.
Thus,  in the case of the SM
we have
\begin{equation}
D_{SM}=(
2h_{W}^{2}+h_{Z}^{2}+2h_{t}^{2})  /8,\ E_{SM}=(  2h_{W}^{3}
+h_{Z}^{3})  /6\pi.
\end{equation}

Since only bosons
contribute to the parameter $E$, extensions of the SM containing extra bosons
are often considered in the literature. However, a cubic term in the effective
potential is difficult to obtain due to the thermal mass $\Pi\sim h^{2}T^{2}$
in Eq.~(\ref{debye}). This has lead to the investigation of scenarios in which
a negative squared mass cancels the thermal mass, $\mu^{2}\simeq-\Pi(T)$. For
high values of the couplings $h_{i}$ the expansion (\ref{fpots}) breaks down.
In this case, the one-loop terms $\phi^{4}\log\phi$ in the zero-temperature
effective potential (\ref{v1loop}) may strengthen the phase transition by
causing a barrier at $T=0$. Two loop contributions and tree-level modifications
to the effective potential have also been considered in order to increase the
strength of the electroweak phase transition. In the present paper we shall
consider all these extensions of the SM.

A simple approximation for the general dependence of the friction coefficient $\eta$ on
the parameters of the model was derived in Refs.~\cite{ariel04,ms10}. We shall use
this approximation to estimate the wall velocity (for a different approach see \cite{hs12}). In this approximation, $\eta$
receives contributions from particles which obey the Boltzmann equation as well as
from infrared excitations of bosonic fields, which are treated classically.
The contribution of Boltzmann particles is given by
\begin{equation}
\eta_{B}=\sum_{i}\frac{g_{i}h_{i}^{4}}{\bar{\Gamma}/T}T\int
_{0}^{\phi_{c}}\left[  c_{1}({m_{i}}/{T})\right]  ^{2}({\phi}/{T})^{2}
\sqrt{2V_{T}}\,d\phi, \label{etath}
\end{equation}
where $\phi_{c}=\phi_{b}(T_{c})$, the function $c_{1}$ is given by
\begin{equation}
c_{1}(x)=\frac{1}{2\pi^{2}}\int_{x}^{\infty}dy\,\sqrt{y^{2}-x^{2}}\frac{e^{y}
}{\left(  e^{y}\mp1\right)  ^{2}},
\end{equation}
and $\bar{\Gamma}$ is an average interaction rate arising from the collision
terms of the Boltzmann equations. The friction decreases with this parameter
since the deviations from the equilibrium distributions in front of the wall
will be smaller if the processes are quicker. For the electroweak phase
transition, $\bar{\Gamma}$ is typically $\sim10^{-2}T$.
The infrared bosons
contribution is given by
\begin{equation}
\eta_{\mathrm{ir}}=\sum_{\mathrm{bosons}}\frac{g_{i}h_{i}^{4}\pi m_{D}^{2}
}{8T^{2}}T\int_{\phi_{0}}^{\phi_{c}}b({m_{i}}/{T})\,({\phi}/{T})^{2}
\sqrt{2V_{T}}\,d\phi, \label{etair}
\end{equation}
where $m_{D}$ is the Debye mass, given by $m_{D}^{2}=(11/6)g^{2}T^{2}$ for the
W and Z bosons of the SM, and $m_{D}^{2}=h^{2}T^{2}/3$ for a scalar singlet.
The integral in (\ref{etair}) has an infrared cut-off $\phi_{0}$ for small
$\mu_{i}$, given by $\phi_{0}=\sqrt{L_{w}^{-2}-\mu_{i}^{2}}/h_i$ for $\mu
_{i}<L_{w}^{-1}$, and $\phi_{0}=0$ for $\mu_{i}>L_{w}^{-1}$, where $L_{w}$ is
the wall width. In the thin wall approximation, $L_{w}$ can be estimated as
$L_{w}\approx\int_{0.1\phi_{c}}^{0.9\phi_{c}}d\phi/\sqrt{2V_{T}}$. The
function $b$ is given by
\begin{equation}
b(  x)  =\frac{1}{2\pi^{2}}\int_{x}^{\infty}\frac{dy}{y^{3}}
\frac{e^{y}}{\left(  e^{y}-1\right)  ^{2}}.
\end{equation}
The two contributions dominate in different parameter regions, and we have
$\eta=\eta_{B}+\eta_{\mathrm{ir}}$.

Depending on the model, we shall use either the full one-loop effective potential (\ref{v1loop}-\ref{f1loop}) or its high-temperature expansion (\ref{fpots}). In the former case, we shall use the friction coefficients (\ref{etath}) and (\ref{etair}), while in the latter case we shall use a similar high-temperature approximation for the friction coefficients,
\begin{eqnarray}
\eta_{B}  &  =&\sum\frac{g_{i}h_{i}^{4}}{\bar{\Gamma}/T}\left(
\frac{\log\chi_{i}}{2\pi^{2}}\right)  ^{2}\frac{\phi^{2}\sigma}{T},
\label{etathap} \\
\eta_{\mathrm{ir}}  &  =&\sum_{\mathrm{bosons}}\frac{g_{i}m_{D}^{2}T}{32\pi
L_{w}}\log\left(  m_{i}(\phi)L_{w}\right)  , \label{etairap}
\end{eqnarray}
and $L_{w}\approx\phi^{2}/\sigma$. Here,  $\chi_{i}=2$ for fermions and $\chi_{i}=m_{i}\left(  \phi\right)  /T$
for bosons, and $\sigma$ is the surface tension of the bubble wall.

\subsection{The SM with a low Higgs mass}

\begin{figure}[bth]
\centering
\epsfysize=6cm \leavevmode \epsfbox{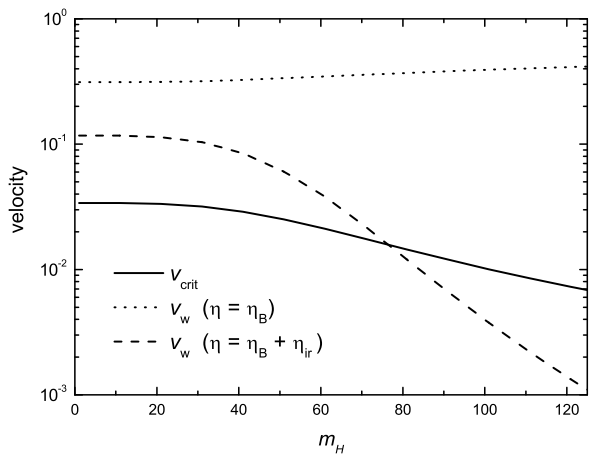}
\caption{The critical velocity $v_{\mathrm{crit}}$ and the wall velocity $v_w$ for the
SM with a low Higgs mass.
}
\label{figsm}
\end{figure}

For comparison with previous results, we  consider first the unrealistic
case of the SM with a light Higgs. We  also consider larger Higgs
masses (although for large $m_{H}$ the
perturbative expansion breaks down)
in order to analyze the dependence of the
velocity on the strength of the phase transition. For this model we use the approximation (\ref{fpots}) for the effective potential as well as the approximations (\ref{etathap}-\ref{etairap}) for the friction.
The result is shown in Fig \ref{figsm}.

The critical velocity  for this model
(solid line) agrees in order of magnitude, but not exactly, with Ref.~\cite{hkllm}. For instance, for $m_{H}=60\mathrm{GeV}$, they obtain
$v_{\mathrm{crit}}\simeq0.035$, while we obtain $v_{\mathrm{crit}}\simeq
0.022$.
As already mentioned,
as a function of the thermodynamic parameters, our result for $v_{\mathrm{crit}}$ is in good agreement with Ref.~\cite{hkllm}. The present discrepancy is due to the rough estimation of the temperature $T_{u}$ for this model in Ref.~\cite{hkllm}.

The dotted line in Fig.~\ref{figsm} corresponds to the wall velocity obtained
from the calculation of the friction using only the Boltzmann equations. We
remark that our approximation  contains a single, effective rate $\bar
{\Gamma}$ instead of the several collision-term parameters coming from the
different interactions. Setting $\bar{\Gamma}=10^{-2}T$, the values of $v_{w}$,
as well as the dependence on $m_{H}$, agree with the results of
Ref.~\cite{mp95b} (and are also close to those of \cite{knr14}). We stress that
the calculations of Refs.~\cite{mp95b,knr14} omit the contribution of infrared
boson excitations to the friction, which can make the wall velocity
significantly smaller \cite{m00}. In the case of the SM, the infrared term
receives contributions  from the gauge bosons. The result for the complete
friction is shown in a dashed line in Fig.~\ref{figsm}. This result agrees with
the estimations of Ref.~\cite{m00}.

We see that the infrared contribution to the friction dominates for weakly
first-order phase transitions, and causes the deflagration to become unstable
(i.e., $v_{w}<v_{\mathrm{crit}}$).

\subsection{Negative squared mass (thermal cubic term)}

The simplest extension of the SM consists of adding one or more gauge-singlet
scalars $S_{i}$ (see, e.g., \cite{cv93,eq93}). In many models, these
bosons constitute a hidden sector which couples only to the SM Higgs doublet
through the so called Higgs portal operator $h_{s}^{2}H^{\dag}H\sum S_{i}^{2}$
(assuming, for simplicity, real fields and universal couplings $h_{i}=h_{s}$).
The scalars may have $SU(2)\times U(1)$-invariant mass terms $\mu_{s}^{2}
S^{2}$ as well as quartic terms $\lambda_{s}S^{4}$. For the moment we shall not consider
the possibility that $S$ develops a VEV.

If $h_{s}\phi_b(T)/T$ is not too large, the free energy is of the form
(\ref{fpots}), with the cubic term replaced by the term (\ref{debye}). The
latter is not as effective as a cubic term in strengthening the phase
transition. The thermal mass is given by $\Pi=(h_{s}^{2}+\lambda_{s})T^{2}/3$
\cite{eq93}. A  negative value of $\mu_{s}^{2}$ may enhance the strength of the
phase transition, since for $\mu_s^{2}\simeq-\Pi(T)$ the term (\ref{debye}) is
effectively of the form $-T\phi^{3}$. This fact is exploited in the case of the
MSSM in the light-stop scenario \cite{cqw96}. Notice that negative values of
$\mu_{s}^{2}$ may induce a nonvanishing expectation value of the extra scalar.
For the case of top squarks, this introduces the danger of color breaking
minima. We shall not take into account this issue here. For the moment we will
just consider the SM plus $g_{s}$ extra bosonic degrees of freedom.  We thus
have an effective potential of the form
\begin{equation}
V_{T}(\phi)=D\left(  T^{2}-T_{0}^{2}\right)  \phi^{2}-TE\phi^{3}-\frac
{g_sT}{12\pi}\left[  h_{s}^{2}\phi^{2}+\frac{h_{s}^{2}+\lambda_{s}}{3}T^{2}
+\mu_{s}^{2}\right]  ^{3/2}+\frac{\lambda}{4}\phi^{4}, \label{potneg}
\end{equation}
with $D=D_{SM}+g_{s}h_{s}^{2}/24$, $E=E_{SM}$,  $T_{0}^{2}=(m_{h}^{2}/4)/D$, and $\lambda=m_H^2/(2v^2)$.

\begin{figure}[bth]
\centering
\epsfysize=6cm \leavevmode \epsfbox{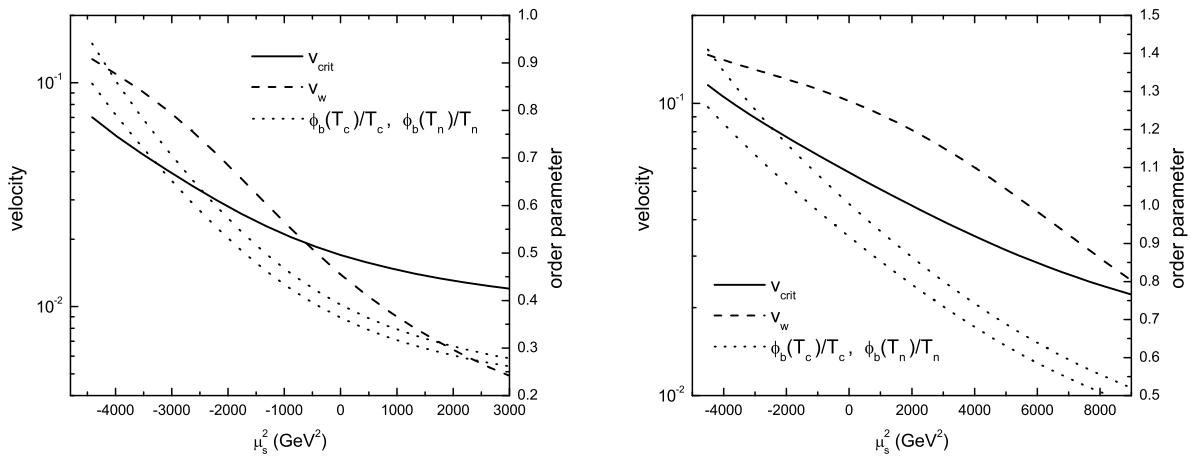}
\caption{The critical velocity $v_{\mathrm{crit}}$, the
wall velocity $v_{w}$, and the order parameter
for an extension of the SM with 6 singlet scalars,
for $\lambda_{s}=0.5$ and $h_{s}=0.7$ (left panel) and $h_{s}=0.8$ (right
panel).
} \label{figmuneg}
\end{figure}

For definiteness, we  consider the case $g_{s}=6$ and $\lambda_{s}=0.5$ (below we consider a different value of $\lambda_s$).
In Fig.~\ref{figmuneg} we show the wall velocity and the critical velocity for a couple of values of
the coupling $h_{s}$, namely, $h_{s}=0.7$ (left panel) and $h_{s}=0.8$ (right
panel). We let $\mu_{s}^{2}$ vary in the range $[-\Pi(T_{0}),\Pi(T_{0})]$ (we
have $T_{c}\simeq T_{0}$). The  dotted curves indicate the value of the order parameter
$\phi_b(T)/T$ at $T=T_{c}$ (lower curve) and at $T=T_{n}$ (upper curve). Since
the plasma is reheated during the expansion and collisions of bubbles, it is
not clear which one is the appropriate value for the baryogenesis condition
(\ref{washout}). The conservative bound is $\phi_b(T_{c})/T_{c}$.
Only for $\mu_{s}^{2}$ close to $-\Pi(T)$ the
contribution of the boson gives a cubic term. Thus, the strength of the phase transition reaches a maximum at this end. Nevertheless, the strength of the
phase transition is also enhanced as $g_{s}$ or $h_{s}$ are increased. For $h_{s}=0.7$ the value $\phi_b/T=1$ is not
reached even in the limit $\mu_{s}^{2}\simeq-\Pi(T)$, while for $h_{s}=0.8$ we have $\phi_b/T\gtrsim1$ for $\mu^{2}<0$.

We have checked that
for different values of $g_{s}$ or $h_{s}$ the behavior of the various curves
is qualitatively similar.
For weakly
first-order phase transitions with small values of $\phi_b/T$ we have a small $v_{w}$ as well as a small $v_{\mathrm{crit}}$, while for more
strongly first-order transitions both $v_{w}$ and $v_{\mathrm{crit}}$ are
higher. On the other hand, their values generally cross at a certain point (as can be appreciated in the left panel of Fig. \ref{figmuneg}).
Thus, we have $v_{w}<v_{\mathrm{crit}}$ for weak enough phase transitions ($\phi_b/T<0.5$) and $v_{w}>v_{\mathrm{crit}}$ for stronger phase transitions. In particular,
for $\phi_b/T\geq1$ we obtain stable deflagrations. Notice, however, that the curves of $v_w$ and $v_{\mathrm{crit}}$ begin to approach each other again as the strength of the phase transition continues increasing. As we shall see below, for very strong phase transitions we will have $v_{w}<v_{\mathrm{crit}}$ again.

For $g_{s}=6$ and  $h_{s}\simeq0.7$ this model can be
regarded as a toy model for the light stop scenario of the MSSM, consisting of the SM plus the light right-handed top squark (stop).
At the one-loop order and
disregarding the possibility that the extra scalar develops a VEV, the main
quantitative difference with the realistic case appears in the resummed daisy
diagrams. The main contribution to the thermal mass of the stop is of the form
$4g_{\mathrm{str}}^{2}T^{2}/9$, where $g_{\mathrm{str}}$ is the strong
gauge coupling \cite{cqw96}. In our numerical calculation, we obtain the same effect by setting the parameter $\lambda_{s}$ in Eq.~(\ref{potneg}) to the value
$4g_{\mathrm{str}}^{2}/3$. This increases considerably the value of
$\Pi(T)$ and, hence, the value of the negative squared mass needed to
compensate this thermal mass.
The collision terms in the
Boltzmann equations are also different due to a different particle content.
This effect  may be relevant if the wall velocity is
close to the critical value.
For the MSSM we can choose the value of our
effective rate $\bar{\Gamma}$ in order to match the result of the detailed
microphysics calculation \cite{js01}. Turning off the infrared contribution to the friction and setting\footnote{In fact, the value of $m_{H}$ does not affect significantly the Boltzmann result, as
observed in the dotted curve of Fig.~\ref{figsm}.}
 $m_{H}=110\mathrm{GeV}$ as in
\cite{js01}, the wall velocity should vary  around  $v_{w}=0.1$ (the friction is higher than in the
SM due to the coupling of the extra boson with the Higgs). We obtain the correct $v_w$ for
$\bar{\Gamma}=5\times
10^{-3}T$.

\begin{figure}[bth]
\centering
\epsfysize=7cm \leavevmode \epsfbox{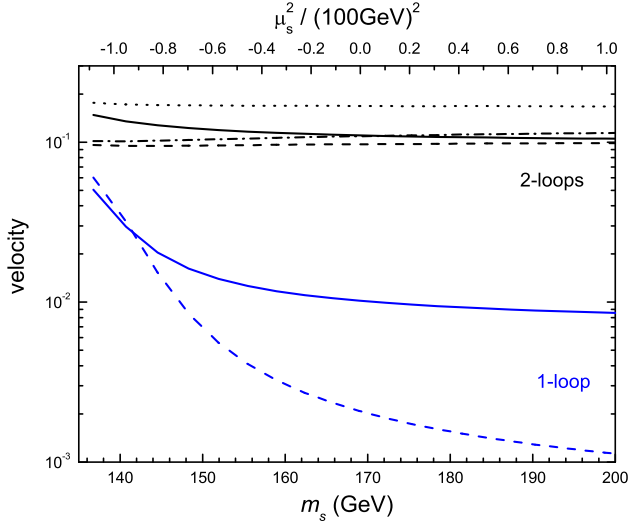}
\caption{The critical velocity (solid lines) and the wall velocity
(dashed lines) for the case of the SM plus a light stop. The two lower (blue) lines correspond
to considering only the 1-loop potential. The upper (black) lines
correspond to including the 2-loop correction. The dash-dotted line corresponds to turning off the contribution of infrared fields. The dotted
line corresponds to increasing the value of the parameter $\bar{\Gamma}$ by a
factor of 2.}
\label{figmssm}
\end{figure}

The lower, blue curves of Fig.~\ref{figmssm} show the
estimated wall velocity (including the infrared part of the friction) together with the critical value for this case. The result is
plotted as a function of $\mu_s^2$ as well as of the mass of the extra scalar, which is given by
$m_{s}=\sqrt{h_{s}^{2}v^{2}+\mu_{s}^{2}}$.
We considered values of $\mu_{s}^{2}$ from
$-\Pi(T_{0})$. For this value of $\mu_s^2$, the
infrared contribution to the friction lowers the wall velocity from $v_{w}\simeq 0.1$ to $v_{w}\simeq 0.05$, and the
effect is stronger for weaker phase transitions (dashed blue curve).

We see that we have $v_w<v_{\mathrm{crit}}$, except for $\mu_s^2$ very close to $-\Pi(T_{0})$. In any case, we obtained values of $\phi_b(T_{c})/T_{c}$ smaller than $0.7$ even in this
limit. Indeed, in this scenario a phase transition which
is strong enough for baryogenesis is obtained only for unrealistic values of
the Higgs mass. The situation improves when two-loop corrections are considered.

\subsection{Two-loop effects (the MSSM)}

Although the light stop scenario does not give a strong enough phase transition at one-loop order, two-loop corrections can make the phase transition strongly first-order even without requiring negative $\mu_s^2$ \cite{e96,cqw98}. The most important two-loop
corrections are of the form $\phi^{2}\log\phi$,
\begin{equation}
V_{2}\left(  \phi,T\right)  \approx C\frac{T^{2}\phi^{2}}{32\pi^{2}}
\log\left(  \frac{\phi}{T}\right)  . \label{v2l}
\end{equation}
We consider such a contribution by adding to the potential (\ref{potneg}) a
term of the form (\ref{v2l}), with a coefficient $C=6h_{s}^{4}
-8g_{\mathrm{strong}}^{2}h_{s}^{2}$ coming from the MSSM stop and gluon loops.

The results for the wall and critical velocities with this modification are
shown in Fig.~\ref{figmssm} (upper, black curves). We obtain  a higher
critical velocity $v_{\mathrm{crit}}\simeq0.1$-$0.15$ (black solid line) as well as a
higher wall velocity $v_{w}\simeq0.1$ (black dashed line), but we have $v_w<v_{\mathrm{crit}}$. Notice that the
strength of the phase transition is essentially due to the presence of the term
(\ref{v2l}) in the effective potential. As a consequence, the result has a very
soft dependence on $m_{s}$.  The significant increase of the wall
velocity with respect to the one-loop value is due to the fact that the two-loop term does not change the particle
content of the model (hence the friction cannot increase significantly) but
does increase the strength of the phase transition. In this approximation we obtain $\phi_b/T\gtrsim1$ for all
the range of $\mu_{s}^{2}$ considered.

Since the wall velocity
is so close to the critical value, the uncertainties in the calculation of
$v_{w}$ become relevant. In this case the Boltzmann part of the friction
dominates. This can be seen by turning off the infrared contribution, which results in the dash-dotted curve of Fig.~\ref{figmssm}.
Since the friction is dominated by the
Boltzmann term, the $\mathcal{O}(1)$ errors in the estimation of collision
terms are consequential for the stability of the deflagration.
To see the sensitivity to the parameter $\bar{\Gamma}$, let us
consider again the SM-like value $\bar{\Gamma}=10^{-2}T$ instead of the MSSM-like value $\bar{\Gamma}=5\times10^{-3}T$. This gives the dotted curve\footnote{In Ref.~\cite{ms10} an even higher value  $\bar{\Gamma}=5\times10^{-2}$ was used for this model, obtaining as a consequence higher values of the wall velocity ($v_{w}\simeq
0.4$-$0.45$).}, which is safe from the instability.
On the other hand, a different approximation for the friction \cite{hs12} (which gives similar values for the one- and two-loop effective potentials) gives $v_{w}\sim0.05$. Such deflagrations would be clearly unstable, since the velocity is a factor of 2 below the critical value.

Therefore, a more accurate determination of  $v_w$ would be important for this model.
It is also important to remark that in baryogenesis calculations the wall velocity is often assumed to be $v_w\lesssim 0.1$, which is just below the lower bound $v_\mathrm{crit}$ for this model. Moreover, this model is severely constrained by experimental data (see, e.g., \cite{cnqw09}).
We stress that we have considered a simplified version of the light stop scenario, which has several parameters we just have not taken into account. In spite of this, we do not expect a significant difference for the critical velocity in the realistic case, although we do expect $\mathcal{O}(1) $ factors  in the wall velocity, as already discussed.

\subsection{Tree-level effects}

Real gauge-singlets allow cubic terms of the form $S^{3}$ or $
H^{\dagger}H  S$. Such corrections to the Lagrangian arise in extensions
of the SM with a singlet (see, e.g., \cite{hjo05,a07,prs07,ak09,ekr12,ck13}) as well as in extensions of the MSSM (see,e.g., \cite{p93,dfm96,hs01,mmw04}). The
presence of cubic terms in the tree-level potential makes it easier to get a
strongly first-order electroweak phase transition. Indeed, the strength of
the transition is dominated by such cubic terms, which provide a barrier
already at zero temperature. In order to
study this effect, one should consider the effective potential for the
condensates of the two fields $H$ and $S$. However, our numerical computation
of the nucleation temperature is based on a single-variable potential. In order
to incorporate this kind of model into our generic analysis, we shall assume
that the thermal tunneling occurs through a trajectory in configuration space
which can be parameterized with a single field $\phi(x)$, and that along this
trajectory the zero-temperature effective potential has a cubic term
\cite{clw13}. This is equivalent to considering a toy model which consists of
adding a term $-A\phi^{3}$ to the tree-level potential (\ref{v0}), where $A$ is
a free parameter with mass dimensions \cite{ms10}. In this model the parameters
of the potential are related to the physical Higgs VEV and mass by
$2m^{2}=\lambda v^{2}-3Av$, $m_{H}^{2}=2\lambda v^{2}-3Av$.

\begin{figure}[bth]
\centering
\epsfysize=6cm \leavevmode \epsfbox{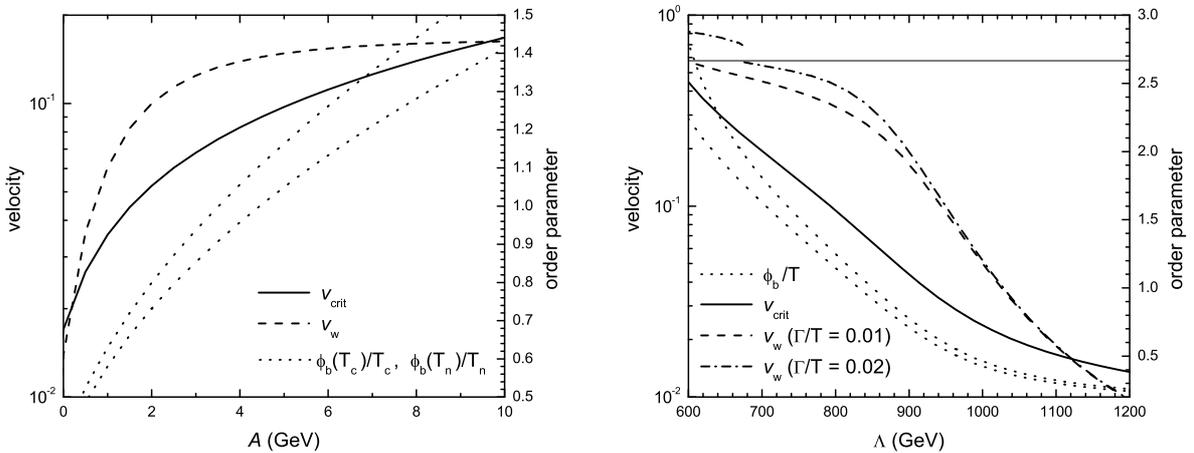}
\caption{
Left panel: the SM plus a singlet, with
a cubic term in the scalar sector. Right panel: the SM plus a sextic operator. The horizontal line indicates the speed of sound.}
\label{figcubic}
\end{figure}

We thus consider the high-temperature effective potential (\ref{potneg}) plus a
term $-A\phi^{3}$. We consider values of the parameters as in the left panel of Fig.~\ref{figmuneg} ($g_{s}=6$, $h_{s}=0.7$, $\lambda_{s}=0.5$,
$\bar{\Gamma}/T=0.01$) with $\mu_{s}=0$. We show the result in the left panel
of Fig.~\ref{figcubic}. We have considered values of the parameter $A$ for
which $h_{s}\phi_b(T_{n})/T_{n}\lesssim1$, so that the high-temperature expansion is valid. The strength of the phase transition grows quickly
with $A$, and values $\phi_b/T>1$ are reached for values of $A$ which are much
smaller than the scale $v$. In this parameter range we obtain subsonic
velocities. These deflagrations are stable on almost the entire considered
range.

Another possible tree-level modification is the introduction of non-renormalizable
operators. In particular, a dimension-six term of the form $(
H^{\dagger}H-v^{2}/2)^{3}/\Lambda^{2}$ gives a negative quartic
coupling if the cutoff is low enough \cite{z93,ho04,gsw05,gt08,dgw08}. We thus consider the SM plus
a term of this form, which gives a term $\left(  \phi^{2} -v^{2}\right)
^{3}/(8\Lambda^{2})$ in the effective potential. Adding this term to the tree-level potential (\ref{v0}) does not shift the
value of the minimum nor the Higgs mass, i.e., we have
$v^{2}=2m^{2}/\lambda$, $m_{H}^{2}=2\lambda v^{2}$. The sextic operator
introduces the terms
\begin{equation}
+\frac{3v^{4}}{8\Lambda^{2}}\phi^{2}-\frac{3v^{2}}{2\Lambda^{2}}\frac{\phi
^{4}}{4}, \label{corrsext}
\end{equation}
which modify the quadratic and quartic terms of the effective potential and, in
particular, may change their sign. Thus, for $\Lambda<\sqrt{3/2}v^{2}
/m_{H}\simeq600\mathrm{GeV}$ the quadratic term becomes positive (already at
zero temperature). Nevertheless, for $\Lambda<\sqrt{3}v^{2}/m_{H}
\simeq840\mathrm{GeV}$ the quartic term becomes negative, causing a barrier in
the zero-temperature effective potential, which is stabilized by the term
$+\phi^{6}/(8\Lambda^{2})$. For this calculation we use the full one-loop
correction (\ref{f1loop}) instead of its high-temperature approximation.

The results for this model are shown in the right panel of Fig.~\ref{figcubic}
as  functions of the cutoff. Below $\Lambda\simeq850\mathrm{GeV}$ we have
$\phi_b(T)/T>1$. Since the particle content is the same as in the SM, we set again $\bar\Gamma=0.01$. The values of the wall velocity (dashed line) are in good agreement with those of Ref.~\cite{knr14} for stronger phase transitions (for instance, for $\Lambda=700$GeV we obtain $v_w=0.45$ while in \cite{knr14} the result is $v_w=0.46$). For weaker phase transitions we obtain smaller values of $v_w$ (e.g., $v_w\simeq 0.17$ for $\Lambda=900$GeV, in contrast to $v_w=0.27$ in \cite{knr14}). This is to be expected since, for weaker phase transitions, the infrared part of the friction becomes noticeable.
Notice that $v_{w}$ is subsonic in all the range of $\Lambda$ we considered, but it approaches the speed of sound
for $\Lambda\simeq600\mathrm{GeV}$. Therefore, an $\mathcal{O}(1)$ variation
of the friction coefficient may give supersonic solutions. To illustrate this
we consider the value $\bar{\Gamma}/T=0.02$ (dash-dotted line). In this case, detonation
solutions appear below $\Lambda\simeq 675\mathrm{GeV}$. Notice also that the curves
for the two values of $\bar{\Gamma}$ coincide for weaker phase transitions,
where the infrared contribution dominates the friction.

Regarding the stability, we see that the wall velocity becomes smaller than the critical value (solid line) only for weak phase transitions with $\phi_b/T<0.3$, corresponding to wall velocities $v_w\lesssim 0.015$.

\subsection{Strong coupling}

Let us consider again the extension of the SM with a singlet scalar. This time
however, instead of considering a negative $\mu_{s}^{2}$ in order to change
the strength of the phase transition, we shall set $\mu_{s}=0$ and vary the
coupling to the Higgs. Therefore, the $\phi$-dependent mass is given by
$m_{s}^{2}(  \phi)  =h_{s}^{2}\phi^{2}$. Besides the cubic term, for high enough $h_{s}$,
the zero-temperature term $m_{s}^{4}\log m_{s}^{2}$ in Eq.~(\ref{v1loop})
becomes relevant. A barrier in the effective potential appears at zero
temperature, and the size of this barrier grows as $h_{s}$ is increased. This
increases the strength of the phase transition and, as a consequence, the
$\phi^{4}\log\phi$ term becomes even more important. In this case we may have large
values of the order parameter $\phi_b/T$ \cite{ms10,eq07}, and we shall not use the high-temperature
expansion.

\begin{figure}[bth]
\centering
\epsfysize=6cm \leavevmode \epsfbox{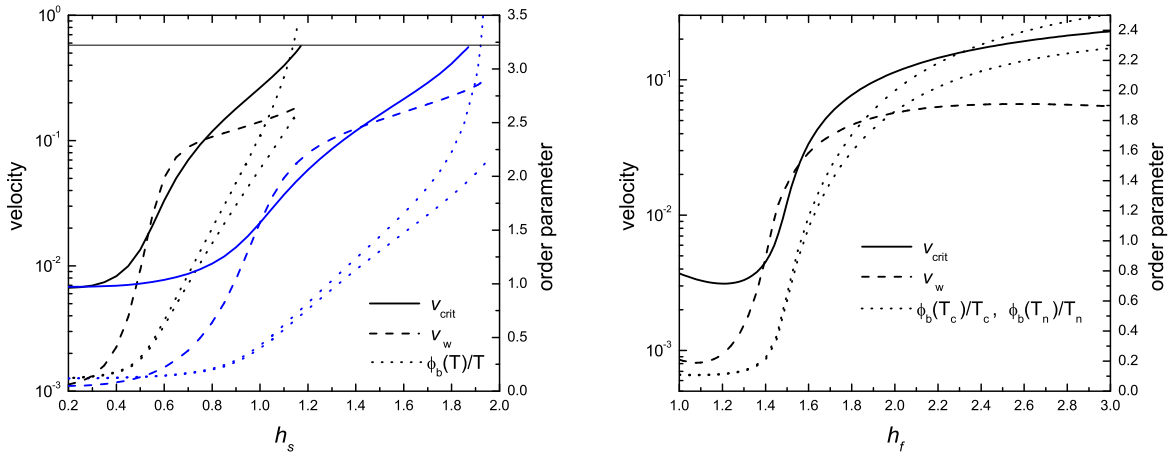}
\caption{The critical velocity, the wall velocity, and the order parameter as
functions of the coupling of the extra particles to the Higgs. Left panel: 12
real scalars (in black) and 2 real scalars (in blue). Right panel: 12
fermionic d.o.f.}
\label{figbf}
\end{figure}

For simplicity, we consider  $\lambda_{s}=0$. The
results are qualitatively similar in the general case (we have checked this)
since $\lambda_{s}$ only affects the value of the thermal mass.
We show the values of the velocity and the order parameter in the left panel of Fig.~\ref{figbf}
as  functions of $h_{s}$ for $g_{s}=2$ bosonic degrees of freedom
(lower, blue curves) and $g_{s}=12$ (upper, black curves). The results are qualitatively very
similar for the two cases, only that for smaller d.o.f.~a higher coupling is
needed to increase the strength of the transition.
In both cases the deflagration
is unstable for weak phase transitions ($\phi_b/T\lesssim0.4$), there is
a range of stable deflagrations, and for stronger phase transitions ($\phi_b/T\gtrsim 1.3$) the velocity is again
smaller than $v_{\mathrm{crit}}$. This tendency is already seen in previous
cases.
There is also
a range of values of $h_{s}$ for which we have both $\phi_b/T>1$ and stable deflagrations. In this
range we have wall velocities  $v_{w}\simeq0.05$-$0.1$, which is good
for baryogenesis.
We obtained subsonic
velocities, but varying the friction by an $\mathcal{O}(1)$ factor detonations
may appear (see, e.g., Ref.~\cite{lms12}).

Extra fermions strongly coupled to the Higgs field can also make the phase
transition strongly first-order \cite{cmqw05}. However, strongly coupled
fermions may make the vacuum unstable due to the minus sign in front of the
term $m_{f}^{4}\log m_{f}^{2}$ in (\ref{v1loop}). This problem can be solved by adding heavy
bosons with the same couplings but with a large $\phi$-independent mass term,
so that they are decoupled from the dynamics at $T\sim100\,\mathrm{GeV}$. A
particular model, considered in Ref.~\cite{cmqw05}, consists in a realization of
split supersymmetry, where the standard relations between the Yukawa and gauge
couplings are not fulfilled. In the simplest case, only $g_{f}=12$ d.o.f.~are
coupled to the SM Higgs, with degenerate eigenvalues of the form $m_{f}
^{2}\left(  \phi\right)  =\mu_{f}^{2}+h_{f}^{2}\phi^{2}$. Here we consider for
simplicity only the case $\mu_{f}=0$. Perturbativity requires $h_{f}
\lesssim3.5$. The bosonic stabilizing fields have the same number of d.o.f.,
and a dispersion relation $m_{b}^{2}\left(  \phi\right)  =\mu_{b}^{2}
+h_{b}^{2}\phi^{2}$, with $h_{b}=h_{f}$ and
\begin{equation}
\mu_{b}^{2}=\left[  \exp\left(  \frac{m_{H}^{2}8\pi^{2}}{g_{f}h_{f}^{4}v^{2}
}\right)  -1\right]  h_{f}^{2}v^{2}, \label{mus}
\end{equation}
which is the maximum value consistent with stability. For simplicity, $\Pi
_{b}=0$ is assumed.

In the right panel of Fig.~\ref{figbf} we show the results for this model as a
function of the coupling $h_{f}$. The velocities, as well as the order
parameter, are smaller than in the case of extra scalars. However, the behavior
is qualitatively similar. In particular, there is a range of values of $h_{f}$
for which the deflagration is stable (i.e., $v_{w}>v_{\mathrm{crit}}$). In this range we have $v_w\sim 10^{-2}$. Values of the order parameter $\phi_b/T\simeq 1$ occur at the upper limit of this range and, thus, may be compatible with stability.

\section{Consequences of the instability} \label{discu}

So far we have only calculated the critical velocity below
which the deflagration is unstable, and compared it with the actual value of the wall velocity.
We now wish to discuss the possible consequences of the case $v_{w}
<v_{\mathrm{crit}}$.

Before doing so, it is worth discussing  on the
use of the  phenomenological form $\eta v\gamma$ for the friction, explicit
in the calculation of $v_w$ and implicit in the calculation of $v_{\mathrm{crit}}$ (or, equivalently, of $\eta_{\mathrm{crit}}$).
Important deviations from this scaling may occur \cite{knr14}. In particular, the friction should saturate in the ultra-relativistic regime \cite{bm09}, while this model grows unboundedly as $v_w\to 1$ due to  the $\gamma$ factor.
We thus may expect significant errors for wall velocities close to the speed of sound or higher. Nevertheless, for small wall velocities (particularly those of interest for baryogenesis) our phenomenological model should give a good description. Indeed, notice that for the two models which admit a direct comparison with the results of \cite{knr14} (namely, the SM with a low Higgs mass and the SM with a low cutoff, for which the particle content is that of the SM), our results for $v_w$ (before taking into account the infrared part of the friction) are in good agreement with those of \cite{knr14}.
In any case, since the scaling $\eta v\gamma$ generally overestimates the friction force, we may expect both $v_w$ and $v_{\mathrm{crit}}$ to be actually higher than our estimations. Therefore, since deflagrations with $v_w<v_{\mathrm{crit}}$ are unstable,  our estimation of $v_{\mathrm{crit}}$ gives a conservative lower bound for the velocity of a stable deflagration.

Let us now assume that we are in the case $v_{w}
<v_{\mathrm{crit}}$. Then,
perturbations of the wall-fluid system in a range of wavenumbers $0<k<k_c$ are unstable. In the first place, it is  important to determine
whether these instabilities will grow in a time comparable to the duration of
the phase transition.
There is a characteristic scale in the problem (see \cite{mm14defla} for  details), which is given by
$d=\sigma/F_{\mathrm{dr}}$, where $\sigma$ is the surface tension and
$F_{\mathrm{dr}}$ is the driving force. Thus, the deflagration wall is unstable for wavelengths
$\lambda$ which are higher than a critical value $\lambda_{c}\propto d$. The
instabilities develop once the bubble reaches a size $R_{b}\sim\lambda_{c}$.
The quantity $d$ is generally of the order of the initial bubble size. In
contrast, the final bubble size $R_{f}$ is in general much higher than $d$. As
a consequence, there will be instabilities in the range $\lambda_{c}
<\lambda<R_{f}$.

To be more precise, in the limit of small supercooling, small
latent heat, and small velocity we have $\lambda_{c}\sim d/[\bar{L}\left(
1-v_{w}/v_{\mathrm{crit}}\right)  ]$. Therefore, unless $v_{w}$ is very close
to $v_{\mathrm{crit}}$, we have $\lambda_{c}\sim d/\bar{L}$. Although $\bar
{L}$ may be quite small and, hence, $\lambda_{c}$ may be a few orders of
magnitude larger than $d$, the final bubble size $R_{f}$ will still be many orders of
magnitude larger \cite{mm14defla}. This is because $R_{f}$ is related to the
cosmological time $H^{-1}$ (where $H$ is the Hubble rate), while $d$ is given
by thermodynamical variables. Roughly, we have $R_{f}/d\sim M_{P}/T$, where
$M_{P}$ is the Planck mass. For the electroweak phase transition we have
$M_{P}/T\sim10^{17}$.

Hence, the instabilities generally begin to grow when bubbles are still very small. On the other hand, the instabilities need time to develop. The growth rate
$\Omega$ in the linear regime is proportional to $\lambda-\lambda_{c}$ and to
$1-v_{w}/v_{\mathrm{crit}}$. A simple dynamical analysis shows that (in the
approximation $T_{n}/T_{c}\simeq1$, $\bar{L}\ll1$, $v_{w}\ll1$) the
instabilities become important when the bubble size reaches the value
$R_{b}^{\mathrm{inst}}\sim\lbrack\bar{L}\left(
1-v_{w}/v_{\mathrm{crit}}\right)  ]^{-2}d$ \cite{mm14defla}. Again, even though
$\bar{L}$ may be quite small we will have $R_{b}^{\mathrm{inst}}\ll R_{f}$
unless $v_{w}$ is extremely close to $v_{\mathrm{crit}}$. Therefore, the
instabilities become important very early in the development of the phase
transition.

As already discussed, the instability of the deflagration may spoil the mechanism of electroweak
baryogenesis if the wall accelerates and reaches supersonic velocities. In fact, the linear stability analysis breaks down as the
perturbations grow. Therefore, $\Omega$ only indicates the initial acceleration rate of
the unstable wall. It is in principle possible that nonlinear effects
stabilize the propagation of the phase transition front. In such a case, one
may expect that the wall reaches a new stationary regime, perhaps with a higher velocity. Notice, however,
that this new regime will not correspond to a stable weak deflagration, unless the conditions which determine $v_w$ change. All other known stationary solutions (namely, Jouguet deflagrations
and weak detonations) are supersonic.

There are (at least) two possible alternatives to this situation. Since the
instabilities corrugate the wall, the growth of the
bubble may be of dendritic type \cite{fa90}. The hydrodynamics in this case may
differ significantly from the case of planar (or spherical) walls, and there
may be new stationary propagation modes. Another possibility is that, by the
time the instabilities begin to become noticeable, the shock fronts coming
from other bubbles already hit the wall. This may happen for small enough
$v_{w}$, since the shock velocity is supersonic. In such a case, the plasma
outside the bubble will be reheated from the initial temperature $T_{n}$ to a
higher temperature $T_{r}$. This reheating will tend to decrease both  $v_{w}$
and  $v_{\mathrm{crit}}$, but the overall effect may be stabilizing. Indeed,
the release of latent heat may reheat the plasma up to $T_{c}$. In such a case
the phase transition necessarily slows down \cite{suhonen,witten,ms08}. The effects of such a reheating on electroweak baryogenesis have been investigated in Refs.~\cite{h95,ariel01,ma05}.

Leaving aside these possibilities, the critical velocity sets a lower bound for the velocity of stationary phase transition fronts.
This lower bound for the wall velocity may be used
to constrain models of EWB in two ways. On the one hand, if the wall
velocity for a given model is calculated and turns out to be below $v_{\mathrm{crit}}$, then in
principle the wall will accelerate to velocities which are too high for
EWB. This would rule out this model as a baryogenesis scenario. On the other hand, since an accurate calculation of
$v_{w}$ is too difficult and model-dependent, using the value of
$v_{\mathrm{crit}}$ as a lower bound may be useful. This lower bound may
constrain the baryogenesis mechanism if $v_{\mathrm{crit}}$ turns out to be too large.

Interestingly, besides the potentially negative consequences for baryogenesis,
the instability of slow walls may give rise to new cosmological consequences
for weak phase transitions. Indeed, the instabilities lead to acceleration of
the walls, anisotropic growth of bubbles and turbulence of the plasma. These
effects may give origin to magnetic fields \cite{soj97} or gravitational waves
\cite{mm14defla}.

\section{Conclusions} \label{conclu}

It is known that the stationary motion of phase transition fronts may be unstable for subsonic
velocities. Specifically, the deflagration front is hydrodynamically unstable
below a critical velocity $v_{\mathrm{crit}}$. In order to discuss the
implications of this fact for the electroweak phase transition, we have
computed both the wall velocity and the critical velocity for several extensions of the Standard Model.  We
have considered a significant variety of models and a wide range of parameters,
including those which are favorable for baryogenesis.

The general result is that the deflagration tends to be unstable for very weak
or very strong phase transitions, while for phase transitions with an order parameter $\phi_b/T\approx 1$ the deflagrations are generally stable.
In general, for any model there is a
range of parameters for which we have stable deflagrations.

The stability condition $v_{w}>v_{\mathrm{crit}}$ constitutes in principle a
restriction for electroweak baryogenesis (assuming that, if it is not
fulfilled, the wall will accelerate to supersonic velocities). This condition, combined
with that of avoiding the washout of the generated BAU (i.e., $\phi_b/T\gtrsim
1$), the requirement of enough CP violation, and experimental bounds, may
restrict significantly some models. We remark that, even if we have a subsonic wall with
$v_w>v_{\mathrm{crit}}$, a departure of the wall velocity from the range
$10^{-2}<v_{w}<10^{-1}$, where the BAU has its maximum, may also prevent a
quantitatively successful EWB. In all the models we
considered we found a region of parameter space for which the deflagration is
stable and the above conditions for $\phi_b/T$ and $v_w$ are fulfilled.

We stress that the friction is very model-dependent, and current calculations have
large errors which propagate to the wall velocity. For  velocities close to the speed of sound, such $\mathcal{O}(1)$
factors may determine whether the wall propagates as a deflagration or a
detonation. Similarly,  for wall velocities close to the critical
value, $\mathcal{O}(1)$ variations of the friction
will determine
the stability or instability of the deflagration. The calculation of the critical velocity is not easy either. We have
used the analytic results of Ref.~\cite{mm14defla}, which were obtained using
several approximations, such as the bag EOS and the assumption of planar
walls. Nevertheless, the value of $v_{\mathrm{crit}}$ is less dependent on
details of the specific model (i.e., it depends only on thermodynamical parameters). In the
appendix we give a simple  fit for $v_{\mathrm{crit}}$ as a function of
the parameters $L/w_{+}(T_{c})$ and $T_{n}/T_{c}$.

We also remark that we have used for our calculations a simple phenomenological model for the friction force. We have argued that this approximation should be good at least for the case of small wall velocities which are required for baryogenesis. The actual value of $v_{\mathrm{crit}}$ is possibly higher than our result. Hence,  our calculation gives only a conservative lower bound for the velocity of a stable deflagration.

\section*{Acknowledgements}

This work was supported by Universidad Nacional de Mar del Plata, Argentina,
grant EXA699/14, and by FONCyT grant PICT 2013 No. 2786.

\appendix{}

\section{Phase transition dynamics} \label{apdyn}

\subsection{Bubble nucleation and expansion}

The bubble nucleation probability per unit volume per unit time is given by
\cite{a81,l83}
\begin{equation}
\Gamma_{n}\left(  T\right)  =A\left(  T\right)  e^{-S_{3}\left(  T\right)
/T}, \label{gamma}
\end{equation}
where $A\left(  T\right)  =\left[  S_{3}\left(  T\right)  /(2\pi T)\right]
^{3/2}$ and $S_{3}(T)$ is obtained by extremizing the three-dimensional
instanton action
\begin{equation}
S_{3}[\phi]=4\pi\int_{0}^{\infty}r^{2}dr\left[  \frac{1}{2}\left(  \frac
{d\phi}{dr}\right)  ^{2}+V_{T}\left(  \phi\left(  r\right)  \right)  \right]
, \label{s3}
\end{equation}
where $V_T(\phi)$ is defined in Eq.~(\ref{vt}). The variation of $S_{3}[\phi]$ gives an equation for the configuration of the
nucleated bubble. The latter is assumed to be spherically symmetric, and its
radial configuration $\phi_{n}(r)$ satisfies the boundary conditions
$d\phi_n/dr|_{r=0}=0,\ \lim_{r\rightarrow\infty}\phi_n\left(  r\right)  =0$.  We  solve the equation for $\phi_{n}(r)$
iteratively by the overshoot-undershoot method (see Ref.~\cite{ms08}
for details).

The nucleation rate vanishes at $T=T_{c}$ and grows rapidly as $T$ descends
below $T_{c}$.  At a temperature $T<T_{c}$, the probability of finding a bubble in  a causal volume $V_{c}\sim(2/H)^{3}$ is given by
\begin{equation}
P(T)=\int_{T_{c}}^{T}\Gamma_{n}(T)V_{c}\frac{dt}{dT}dT. \label{intnucl}
\end{equation}
The
time-temperature relation is given by $dT/dt=-HT$, where the expansion rate is given by
$H=\sqrt{8\pi G\rho_{u}(T)/3}$, with $G$ the Newton's constant.
We define as usual the nucleation temperature $T_{n}$ at which nucleation effectively begins  by the condition
$P(T_{n})=1$. We  compute $T_{n}$ numerically from Eqs.~(\ref{gamma}
-\ref{intnucl}).

\subsection{Hydrodynamics}

The equation for the wall can be obtained from the phenomenological equation for $\phi$, which is similar to Eq. (\ref{eqphi}), with the last term replaced by a phenomenological term proportional to $ u^{\mu}\partial_\mu\phi$. The usual procedure is to consider the equation in the reference frame of the planar wall, multiply by $\partial \phi/\partial z$, where $z$ is the coordinate perpendicular to the wall, and then integrate across the wall, taking into account the variation of the fluid variables. To perform the integration, either an ansatz for the wall profile is used or some approximations are needed. We shall use the result of \cite{mm14defla}, which uses linear
approximations for the variations of  quantities inside the wall. We have
\begin{equation}
p_{u}(T_{u})-p_{b}(T_{b})-\left\langle \frac{dp}{dT^{2}}\right\rangle \left(
T_{u}^{2}-T_{b}^{2}\right)  -\eta\left\langle \gamma v\right\rangle =0.
\label{eqmicro}
\end{equation}

We also use the bag EOS,
\begin{equation}
p_{u}\left(  T\right)  =\frac{a}{3}T^{4}-\frac{L}{4},\quad p_{b}\left(
T\right)  =\left(  \frac{a}{3}-\frac{L}{4T_{c}^{4}}\right)  T^{4}, \label{eos}
\end{equation}
where $L$ is the latent heat and $a$ is an effective radiation constant
depending on the number of relativistic d.o.f.  For the bag EOS, Eq.~(\ref{eqmicro}) becomes
\begin{equation}
\frac{L}{4}\left(  1-\frac{T_{u}^{2}T_{b}^{2}}{T_{c}^{4}}\right)
=\eta\left\langle \gamma v\right\rangle . \label{forcebag2}
\end{equation}
The temperature $T_{b}$ behind the wall is related to the temperature
$T_{u}$ in front of it through the equations for the discontinuity, Eqs.~(\ref{eqlandau}). For the bag EOS we have
\begin{equation}
T_{b}^{4}={\frac{v_{u}\gamma_{u}^{2}}{v_{b}\gamma_{b}^{2}(1-\bar{L})}
}T_{u}^{4}, \label{tbtu}
\end{equation}
where
$\bar{L}={L}/{(4aT_{c}^{4}/3)}={L}/{w_{u}(T_{c})}$. 
The velocities $v_u$ and $v_b$ are related by
\cite{s82}
\begin{equation}
v_{u}=\frac{1}{1+\alpha_{u}}\left[  \frac{1}{6v_{b}}+\frac{v_{b}}{2}\pm
\sqrt{\left(  \frac{1}{6v_{b}}+\frac{v_{b}}{2}\right)  ^{2}+\alpha_{u}
^{2}+\frac{2}{3}\alpha_{u}-\frac{1}{3}}\right]  , \label{steinhardt}
\end{equation}
with
$\alpha_{u}=L/\left(  4aT_{u}^{4}\right)  =(\bar{L}/3)(T_{c}/T_{u})^{4}$.

We see that we have two kinds of solutions. Indeed, the $+$ sign in front of the square root in Eq.~(\ref{steinhardt}) corresponds to detonations while the
$-$ sign corresponds to deflagrations. For detonations we have $v_{b}<v_{u}$
while for deflagrations we have $v_{b}>v_{u}$. These solutions are further
classified into weak, strong or Jouguet, depending on the relation of the
outgoing velocity $v_{b}$ with the speed of sound $c_{s}=1/\sqrt{3}$. For weak deflagrations we have $v_b<c_s$.

The fluid profiles away from the wall are quite simple for
a planar interface. For a weak deflagration we can only have constant fluid velocities. In the reference frame of
the bubble center, the  velocity profile  is the following. Inside the bubble the fluid is at rest (in the reference frame of the wall, this gives the condition $v_{b}=v_{w}$). In front of the wall we have a constant fluid velocity
$\tilde{v}_{u}>0$ (in the wall frame, this inequality transforms into the
deflagration relation $v_{u}<v_{b}$). The boundary condition of an unperturbed fluid far in front of the
wall is fulfilled by a jump of the fluid velocity from $\tilde{v}_{u}$ to $0$ at a
certain point. Such a discontinuity in the fluid profile is called a shock
front.

The shock discontinuity is also determined by applying
Eqs.~(\ref{eqlandau}) to the shock front, where the EOS is now the same in both sides of
the interface. It turns out that the shock front moves supersonically. The
region between the bubble wall and the shock front is reheated, i.e., the
temperature $T_{u}$ is higher than the nucleation temperature $T_{n}$. We
have
\begin{equation}
\frac{\sqrt{3}\left(  T_{u}^{4}-T_{n}^{4}\right)  }{\sqrt{\left(  3T_{u}
^{4}+T_{n}^{4}\right)  \left(  3T_{n}^{4}+T_{u}^{4}\right)  }}=\frac
{v_{b}-v_{u}}{1-v_{u}v_{b}}. \label{shock}
\end{equation}
Using the relations (\ref{tbtu}-\ref{steinhardt}), the two equations
(\ref{forcebag2}) and (\ref{shock}) can be readily solved to obtain $v_{w}$ as
a function of $\eta$ and $T_{n}$.

\section{The critical velocity} \label{apvcrit}

\begin{figure}[bth]
\centering
\epsfysize=6cm \leavevmode \epsfbox{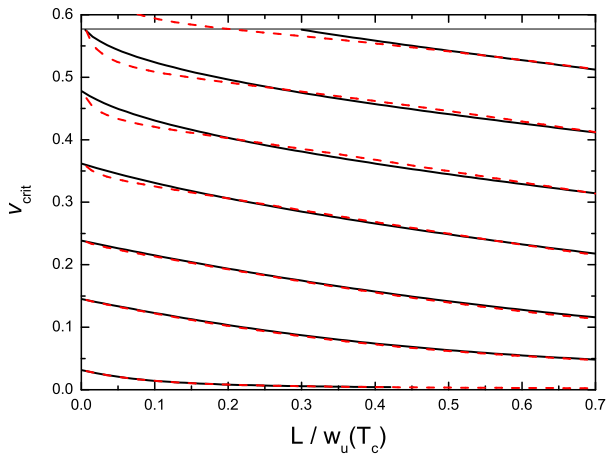}
\caption{The critical velocity (solid lines) and the fit (dashed
lines) as functions of the parameter $\bar{L}$ for several amounts of
supercooling. From bottom to top we have $T_{n}/T_{c}=0.999$, $0.98$, $0.95$,
$0.9$, $0.85$, $0.8$ and $0.75$. The horizontal line indicates the value
$c_{s}=1/\sqrt{3}$.}
\label{figvcrit}
\end{figure}

Using the relations (\ref{tbtu}-\ref{shock}) in Eq.~(\ref{eqvcrit}), we obtain the
equation for  the critical velocity
$v_{\mathrm{crit}}$ as a function of $T_{n}/T_{c}$ and $\bar{L}$. The solution
for $v_{\mathrm{crit}}$ is plotted in Fig.~\ref{figvcrit} (solid lines) as a
function of $\bar{L}$ for different values of $T_{n}/T_{c}$. Technically,
solving Eq.~(\ref{eqvcrit}) becomes difficult in some cases due to the
divergence of the quantity $\beta$ as the product $T_{u}T_{b}$ approaches
$T_{c}^{2}$. This happens for instance for high values of the latent heat,
since the reheating in front of the wall may cause $T_{u}$ to exceed $T_{c}$.
As can be seen from Eq.~(\ref{forcebag2}), the exact limit cannot be reached and
we always have $T_{u}T_{b}<T_{c}^{2}$ . As a consequence, there is always a
critical velocity.

In the case of small supercooling and
small latent heat, the wall velocity is generally small, since we have $\eta v_{w}\simeq L(1-T_{u}/T_{c})$.
In this limit we also
have $T_{b}\simeq T_{u}$ and $v_b\simeq v_u$, and
we have a simple expression for $v_{\mathrm{crit}}$ as a function of $T_u$ \cite{mm14defla},
\begin{equation}
v_{\mathrm{crit}}^{2}\simeq(1+\bar{L})\left(  1-\frac{T_{u}}{T_{c}}\right)  .
\label{vcraptu}
\end{equation}
On the other hand,  the temperature $T_u$ of the reheated plasma in front of the wall depends on $v_w$. In the present
approximation we have $T_{u}
/T_{c}\simeq T_{n}/T_{c}+(\bar{L}/\sqrt{3}v_{w})$. Hence, for $v_{w}
=v_{\mathrm{crit}}$  Eq.~(\ref{vcraptu}) gives a quadratic equation. The solution is
\begin{equation}
v_{\mathrm{crit}}\simeq\frac{\bar{L}(1+\bar{L})}{2\sqrt{3}}\left[
\sqrt{1+\frac{12}{\bar{L}^{2}(1+\bar{L})}\left(  1-\frac{T_{n}}{T_{c}}\right)
}-1\right]  . \label{vcraptn}
\end{equation}
This simple approximation
underestimates the value of the critical velocity. It becomes exact in the limit
$T_{n}\rightarrow T_{c}$, but
only gives a good estimation for small amounts of supercooling. For instance, for $T_{n} /T_{c}>0.98$
the error is less than a 5\%, while for $T_{n}/T_{c}=0.9$ the actual value of the critical velocity is about a 15\%
higher than the approximation (\ref{vcraptn}).

Nevertheless, a slight modification of Eq.~(\ref{vcraptn}) provides a good fit,
\begin{equation}
v_{\mathrm{crit}}\simeq\frac{\bar{L}(1+\bar{L})}{2\sqrt{3}}\left[
\sqrt{1+\frac{12}{\bar{L}^{2+\varepsilon}(1+a\bar{L})}\left(  1-\frac{T_{n}
}{T_{c}}\right)  }-1\right]  ,
\end{equation}
with $\varepsilon=0.5(1-T_{n}/T_{c})$ and $a=1-2.7(1-T_{n}/T_{c})$. The error
of this approximation is less than a 5\%
for $0.75<T_{n}/T_{c}<1$
and in all the range $0<\bar{L}<1$ (see Fig.~\ref{figvcrit}).

\end{document}